\documentclass[sigconf, nonacm]{acmart}

\AtBeginDocument{%
  }

\usepackage{xspace}
\usepackage{enumitem}
\usepackage{multirow}

\DeclareMathOperator*{\argmin}{argmin}

\begin{document}

\title{Hologram Representation via Quadratic Phase Gaussian Splatting}

\author{Haolong Wang}
\orcid{0009-0003-5174-9230}
\affiliation{%
  \institution{Swansea University}
  \city{Swansea}
  \country{United Kingdom}}
\email{haolong.wang.00@gmail.com}

\author{Yicheng Zhan}
\orcid{0009-0006-5936-1929}
\affiliation{%
  \institution{University College London}
  \city{London}
  \country{United Kingdom}}
\email{yicheng.zhan.21@ucl.ac.uk}

\author{Kaan Ak\c{s}it}
\orcid{0000-0002-5934-5500}
\affiliation{%
  \institution{University College London}
  \city{London}
  \country{United Kingdom}}
\email{kaanaksit@kaanaksit.com}

\author{Simeng Qiu}
\orcid{0000-0002-0809-0093}
\affiliation{%
  \institution{Swansea University}
  \city{Swansea}
  \country{United Kingdom}}
\email{simeng.qiu@swansea.ac.uk}

\begin{abstract}
We introduce Complex-Valued Quadratic Phase Gaussian (CVQPG), a novel hologram representation method that augments each 2D Gaussian primitive with a quadratic phase profile controlled by a learnable curvature parameter. Against the planar Gaussian baseline, CVQPG improves the average PSNR of holographic reconstructions by $0.19$\,dB (RGB) and $0.33$\,dB (grayscale) at equal primitive counts, and by $0.05$\,dB (RGB) and $0.08$\,dB (grayscale) at equal parameter counts, where it still leads in all visual quality metrics. Our frequency-domain analysis shows that CVQPG better preserves the mid-to-high frequency band of natural images, where the reconstruction MSE drops by up to $11\%$ (RGB) and $22\%$ (grayscale), indicating that modulating primitive wavefronts is an effective and lightweight enhancement. 
\end{abstract}



\maketitle

\section{Introduction}
Computer-Generated Holography (CGH) plays an essential role in applications such as 3D displays~\cite{10108460, zhan2025complexvalued2d, Zhang:17}. Given the dense spatial variations in holograms, designing efficient representations that preserve high-frequency details remains a major challenge in CGH. Implicit Neural Representations (INR)~\cite{sitzmann2020implicitneuralrepresentationsperiodic} favor low-frequency content, and Gaussian image representations~\cite{zhang2024gaussianimage1000fpsimage} fit 2D Gaussians to pixels for compression. A separate line of research tackles hologram novel view synthesis using Gaussians~\cite{10.1145/3731163, 10.1145/3804450}. Closest to us, \citet{zhan2025complexvalued2d} place complex-valued Gaussians directly on the 2D hologram plane to model interference and diffraction, enabling efficient and high-fidelity reconstruction. 

The work by \citet{zhan2025complexvalued2d} assumes a flat phase profile for each primitive, leaving the representational capacity of Gaussian primitives an under-explored question. We further explore this topic by representing full-complex holograms using primitives that explicitly model the curvature of the field. Our proposed Complex-Valued Quadratic Phase Gaussian (CVQPG) augments each Gaussian with a quadratic phase profile controlled by a learnable curvature parameter. We also derive a per-primitive curvature bound from the display's Nyquist sampling limit, ensuring robust hologram reconstruction without high-frequency aliasing.

Our experiments show that CVQPG achieves consistent quality improvements and outperforms the baseline in parameter effectiveness. We investigate the frequency domain features of CVQPG and ablate its components. Our code is available at \href{https://github.com/gsmark36/complex-valued-quadratic-phase-gaussian}{\textbf{GitHub:gsmark36/ complex-valued-quadratic-phase-gaussian}}. 

\section{Methods}
\paragraph{Problem Definition}
The holographic display can reconstruct a 3D scene as multi-plane intensity images by propagating light from a complex hologram $\mathbf{H} \in \mathbb{C}^{N_c \times H \times W}$ with $N_c$ color channels. To synthesize the intensity target for each depth plane, we take an image $\mathbf{I}_{target} \in \mathbb{R}^{N_c \times H \times W}$ and its depth map $\mathbf{D} \in \mathbb{R}^{H \times W}$ as the input for the target synthesis function $f^{m}_{synth}$ at plane $m$. The optimization problem is therefore formulated as:
\begin{equation}
\hat{\mathbf{H}} \leftarrow \argmin_{\mathbf{H}}\ \sum_{m=1}^{M} \mathcal{L} \big( f^m_{recon}(\mathbf{H}), \ f^m_{synth}(\mathbf{I}_{target}, \mathbf{D}) \big),
\label{eq:hologram_optimization}
\end{equation}
where $M$ denotes the total number of depth planes, $f^{m}_{recon}$ is the reconstruction function at plane $m$, and $\mathcal{L}$ represents the loss function that measures the difference between the reconstructed image and the synthesized target. 

\begin{figure*}[t]
\centering
\includegraphics[width=\textwidth]{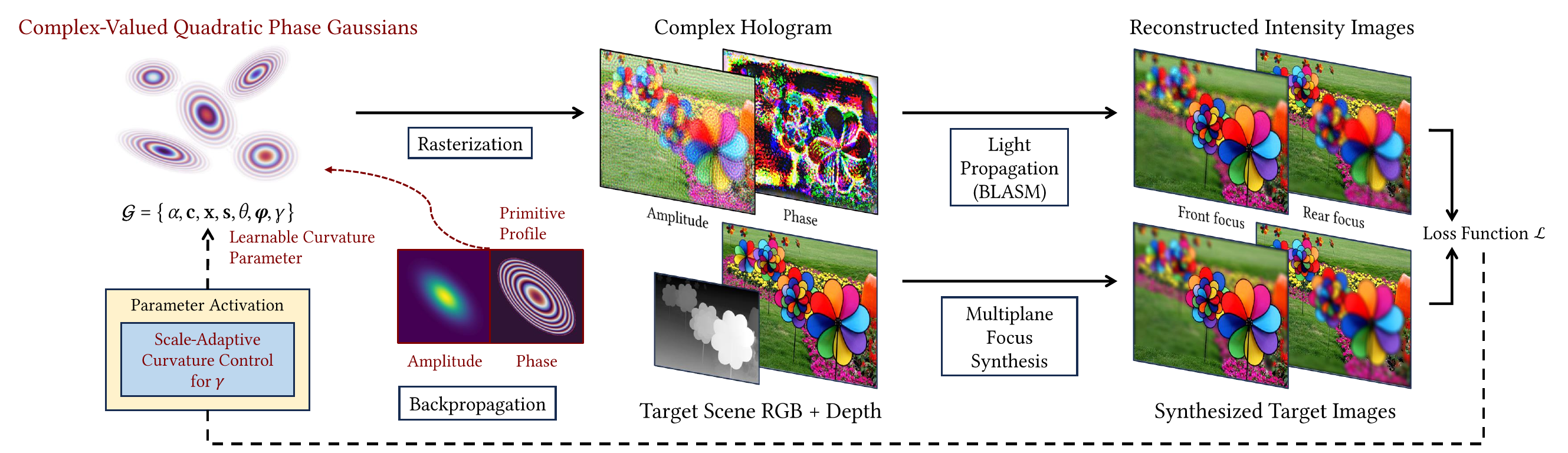}
\caption{
    CVQPG represents full complex holograms with an extra learnable curvature parameter $\gamma$, and optimizes the hologram by comparing the reconstructed intensity images with the synthesized targets. (Source Image: \cite{windmill2013}) 
}
\Description{
    A left-to-right pipeline diagram. On the left, Gaussian primitives are labeled with the parameter set G = alpha, c, x, s, theta, phi, gamma. A rasterization arrow leads to a complex hologram shown as separate amplitude and phase panels. A light propagation arrow leads to reconstructed intensity images at a front and a rear focal plane, which feed a loss function L on the right. A lower branch shows a target RGB image and its depth map passing into two synthesized target images that also feed the loss. A backward-process arrow returns from the loss a parameter activation box that contains a highlighted scale-adaptive curvature control box, whose output updates the Gaussian primitive parameters. 
}
\label{fig:pipeline}
\end{figure*}

\paragraph{Complex-Valued Planar Gaussian}
Our method builds upon the complex-valued 2D Gaussian representation~\cite{zhan2025complexvalued2d}, which we refer to as the planar Gaussian baseline model. It encodes a hologram as a set of $N$ Gaussian primitives. Each primitive is parameterized as $\mathcal{G} = \{\alpha, \mathbf{c}, \mathbf{x}, \mathbf{s}, \theta, \boldsymbol{\varphi}\}$, where $\alpha \in \mathbb{R}$ denotes the opacity, $\mathbf{c} \in \mathbb{R}^{N_c}$ the per-channel color amplitude, $\mathbf{x} \in \mathbb{R}^2$ the 2D position, $\mathbf{s} \in \mathbb{R}^2$ the scales, $\theta \in \mathbb{R}$ the rotation angle, and $\boldsymbol{\varphi} \in \mathbb{R}^{N_c}$ the per-channel phase. The distribution of the Gaussian primitive is determined by the 2D covariance matrix $\boldsymbol{\Sigma} = \mathbf{R} \mathbf{S} \mathbf{S}^\top \mathbf{R}^\top$, where $\mathbf{R}$ is the rotation matrix of angle $\theta$, and $\mathbf{S} = \operatorname{diag}(\mathbf{s})$ is the scaling matrix of scales $\mathbf{s}$. Therefore, the contribution of a Gaussian primitive at pixel coordinate $\mathbf{p}$ is
\begin{equation}
\mathcal{G}(\mathbf{p}) = \alpha \cdot \mathbf{c} \cdot \exp\!\big( -\tfrac{1}{2} (\mathbf{p} - \mathbf{x})^\top \boldsymbol{\Sigma}^{-1} (\mathbf{p} - \mathbf{x}) \big) \cdot \exp\!\big( j \boldsymbol{\varphi} \big).
\label{eq:cvg}
\end{equation}
The phase profile of each primitive is flat across its effective area, equivalent to a plane wave emitted from a Gaussian ellipse. 

\begin{figure}[h]
    \centering
    \includegraphics[width=0.85\columnwidth]{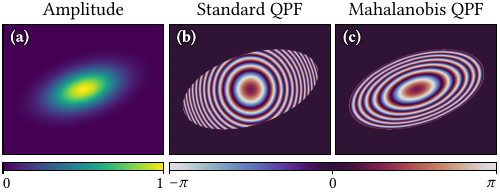}
    \caption{
        Comparison of QPFs on a Gaussian primitive. (a) Shared amplitude envelope. Unlike (b) the standard QPF with misaligned circular fringes, (c) the Mahalanobis QPF yields elliptical fringes perfectly aligned with the envelope. 
    }
    \Description{
        Three square panels in a row. Panel (a) displays an elliptical amplitude heatmap. Panels (b) and (c) display wrapped phase maps: the fringes form concentric circles in (b), but form concentric ellipses matching the envelope's shape in (c). Color scales for amplitude and phase are positioned below. 
    }
    \label{fig:unit_test}
\end{figure}

\paragraph{Complex-Valued Quadratic Phase Gaussian}
Considering the representational capacity, we present a novel primitive: Complex-Valued Quadratic Phase Gaussian (CVQPG). Our design incorporates the Quadratic Phase Factor (QPF) to explicitly modulate field curvature with a minimal overhead of one extra parameter. Conventionally, we modulate a Gaussian primitive with a standard QPF,
\begin{equation}
Q_{std}(\mathbf{p}) = \exp\!\big( j \frac{k}{2f} \| \mathbf{p} - \mathbf{x} \|^2 \big),
\label{eq:qpf1}
\end{equation}
where $k$ is the wavenumber, and $f$ is the focal length. However, the standard QPF models an isotropic circular wavefront, leading to a mismatch with the Gaussian's anisotropic amplitude envelope (Fig.~\ref{fig:unit_test}). To resolve this, we utilize the squared Mahalanobis distance $D^2_{mah}(\mathbf{p}) = (\mathbf{p} - \mathbf{x})^\top \boldsymbol{\Sigma}^{-1} (\mathbf{p} - \mathbf{x})$ as a unified spatial metric. Therefore, the shape-adaptive Mahalanobis QPF is formulated as
\begin{equation}
Q_{mah}(\mathbf{p}) = \exp\!\big( j \frac{k \gamma \det(\mathbf{S})}{2} D^2_{mah}(\mathbf{p}) \big),
\label{eq:qpf2}
\end{equation}
where $\gamma = 1/f$ is the learnable curvature parameter representing optical power, and $\det(\mathbf{S})$ is the determinant of the scaling matrix $\mathbf{S}$. Consequently, each CVQPG primitive is parameterized as $\mathcal{G} = \{\alpha, \mathbf{c}, \mathbf{x}, \mathbf{s}, \theta, \boldsymbol{\varphi}, \gamma\}$, with an added scalar parameter $\gamma \in \mathbb{R}$. Hence, the contribution of a CVQPG primitive at pixel coordinate $\mathbf{p}$ is given by
\begin{equation}
\mathcal{G}(\mathbf{p}) = \alpha \cdot \mathbf{c} \cdot \exp\!\big( -\tfrac{1}{2} D^2_{mah}(\mathbf{p}) \big) \cdot \exp\!\big( j \boldsymbol{\varphi} \big) \cdot Q_{mah}(\mathbf{p}).
\label{eq:cvqpg}
\end{equation}
The configuration of a CVQPG primitive is equivalent to the combination of a planar Gaussian light source and a virtual lens with optical power $\gamma$. During optimization, we symmetrically bound $\gamma$ within $[-\gamma_{\max}, \gamma_{\max}]$, which enables a continuous wavefront transformation from divergence ($\gamma > 0$) to convergence ($\gamma < 0$). We can represent a planar wavefront by $\gamma = 0$, avoiding numerical instability caused by the focal length $f \to \pm \infty$. 


\paragraph{Scale-Adaptive Curvature Control}
The quadratic phase profile acts as a linear chirp signal, providing high-frequency components for detailed hologram reconstruction. However, to prevent aliasing artifacts caused by the physical bandwidth limitations of the Spatial Light Modulator (SLM), we impose a frequency constraint on each primitive. By restricting the chirp evaluation to the $3\sigma$ footprint of each primitive, the maximum instantaneous spatial frequency is 
\begin{equation}
\nu_{\max} = \frac{\gamma \cdot 3\sigma \cdot (\Delta x)^2}{\lambda},
\label{eq:max_freq}
\end{equation}
where $\sigma = \max(\mathbf{s})$ is the equivalent standard deviation, $\lambda = 2\pi / k$ is the wavelength, and $\Delta x$ is the pixel pitch of the SLM. Constraining this peak frequency below the SLM's Nyquist limit ($\nu_{\max} \leq 0.5$ cycles/pixel) directly yields the scale-adaptive curvature bound for each primitive:
\begin{equation}
\gamma_{\max} = \eta \cdot \frac{\lambda}{6\sigma \cdot (\Delta x)^2},
\label{eq:max_curv}
\end{equation}
where $\eta = 0.9$ is a relaxation factor. 

\paragraph{Hologram Rendering and Optimization}
In the rendering process, rasterization is applied to accumulate all primitive contributions (Eq.~\ref{eq:cvqpg}) across the pixel grid to form the hologram,
\begin{equation}
\mathbf{H} = \Bigl\{ \sum_{n=1}^{N} \mathcal{G}_n(\mathbf{p}) \;\Big|\; \mathbf{p} \in [1,W] \times [1,H] \Bigr\},
\label{eq:hologram}
\end{equation}
where $H \times W$ is the size of the hologram. We propagate the hologram via the conventional Band-limited Angular Spectrum Method (BLASM) to reconstruct the field intensity $I_m = \big| \operatorname{BLASM}_m (\mathbf{H}) \big|^2$ at plane $m$. At each plane, we compare the reconstructed intensity $I_m$ with the synthesized target intensity $\hat{I}_m$, minimizing the visual difference between them. The per-plane loss function consists of a reconstruction loss $\mathcal{L}_{recon}$ proposed by \citet{10108460} and a similarity loss $\mathcal{L}_{SSIM} = (1-\operatorname{SSIM})$. To accurately simulate depth-dependent focus and defocus effects, the total training loss $\mathcal{L}$ is summed over all depth planes as
\begin{equation}
\mathcal{L} = \sum_{m=1}^{M} \big( \mathcal{L}^m_{recon}(I_m, \hat{I}_m) + w_1 \cdot \mathcal{L}^m_{SSIM}(I_m, \hat{I}_m) \big),
\label{eq:total_loss}
\end{equation}
where $w_1 = 0.005$ is the SSIM loss weight. 

\paragraph{Curvature Warm-Up Strategy}
We optimize the learnable curvature parameter $\gamma$ via a two-stage strategy. We initialize and lock the curvature at $\gamma = 0$ for all primitives, utilizing the planar Gaussians to optimize the low-frequency patterns. In the refinement stage, the curvature parameter is unlocked and optimized with other parameters. This coarse-to-fine approach stabilizes early-stage training and allows the model to refine the residual high-frequency details. 

\section{Evaluation and Discussion}
\paragraph{Implementation}
We utilize a $3.74\,\mu\text{m}$ pixel pitch and RGB wavelengths of $639\,\text{nm}$, $532\,\text{nm}$, and $473\,\text{nm}$ to propagate the holograms to two depth planes at $1\,\text{mm}$ and $5\,\text{mm}$. Shared parameters follow the same initialization schemes as the baseline model~\cite{zhan2025complexvalued2d}. $\gamma_{\max}$ is calculated from the shortest wavelength, and $\gamma$ is bounded by a $\tanh$ activation. We train the models using identical PyTorch pipelines and the Adan optimizer~\cite{10586270}. The training runs for 2000 steps, with a 400-step warm-up stage. All experiments are conducted on a single NVIDIA TITAN RTX GPU. We display the reconstructed holograms on our holographic display prototype (see Supplementary Section~1). 

\begin{figure}[h]
    \centering
    \includegraphics[width=0.9\columnwidth]{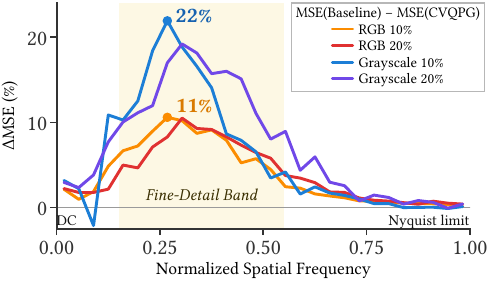}
    \caption{
        Frequency-resolved $\Delta$MSE of CVQPG over the baseline for ten scenes. The gains cover the mid-to-high frequency band that contains fine details of natural images. 
    }
    \Description{
        A line plot of percentage error reduction versus normalized spatial frequency, with four curves for the RGB and grayscale settings at $10\%$ and $20\%$ budgets. All curves start near zero at low frequency, rise to a peak in the mid-to-high frequency band, and decay back toward zero near the Nyquist frequency. the grayscale curves peak higher than the RGB curves. 
    }
    \label{fig:frequency}
\end{figure}

\paragraph{Quantitative Evaluations}
The visual quality of reconstructed images is evaluated using five metrics, including Peak Signal-to-Noise Ratio (PSNR), Structural Similarity (SSIM), Perceptual Similarity Metric (LPIPS)~\cite{8578166}, FLIP~\cite{10.1145/3406183}, and ColorVideoVDP (CVVDP)~\cite{10.1145/3658144}. Our dataset is derived from both DIV2K~\cite{8014884} and the Real Forward-Facing dataset~\cite{10.1145/3306346.3322980}. Then we generate the depth map using Depth Anything V2~\cite{yang2024depthv2}. We first compare CVQPG against the planar Gaussian baseline on ten scenes at a resolution of $640 \times 480$ with compression ratios of $10\%$ and $20\%$. Under equal primitive counts, CVQPG outperforms the baseline across all metrics, achieving mean PSNR gains of $+0.19$\,dB in RGB and $+0.33$\,dB in grayscale (Tab.~\ref{tab:main_results}). Since CVQPG adds one trainable scalar per primitive that brings minor parameter overhead, we further conduct an equal-parameter evaluation by enlarging the baseline's primitive count by $13/12\times$ in RGB and $9/8\times$ in grayscale (see Baseline$^{+}$ rows in Tab.~\ref{tab:main_results}). Per-scene results are given in Section~3 of the supplementary material. To assess improvements in the frequency domain, we measure the reconstruction MSE at each spatial frequency against the target. Evaluated on ten scenes at an equal primitive count (Fig.~\ref{fig:frequency}), the MSE reductions of CVQPG over the baseline are negligible near zero frequency, peak in the mid-to-high frequency band ($11\%$ RGB, $22\%$ grayscale), and vanish near the Nyquist limit. Training time and peak GPU memory are reported in Section~2 of the supplementary material. 

\begin{table}[h]
\centering
\caption{
    Quantitative comparison averaged over ten scenes and two depth planes. We compare CVQPG against the standard Baseline (equal primitive count $N$) and the enlarged Baseline$^{+}$ (equal parameter count). $\Delta$ rows show CVQPG's improvement over the standard Baseline. Best in \textbf{bold}. 
}
\label{tab:main_results}
\resizebox{\linewidth}{!}{%
\begin{tabular}{llcccccc}
\toprule
Setting & Method & $N$ & PSNR$\uparrow$ & SSIM$\uparrow$ & LPIPS$\downarrow$ & FLIP$\downarrow$ & CVVDP$\uparrow$ \\
\midrule
\multirow{4}{*}{RGB $20\%$}
 & Baseline & 30{,}720 & 29.265 & 0.8307 & 0.2895 & 0.1144 & 9.168 \\
 & Baseline$^{+}$ & 33{,}280 & 29.386 & 0.8336 & 0.2853 & 0.1134 & 9.186 \\
 & CVQPG & 30{,}720 & \textbf{29.450} & \textbf{0.8364} & \textbf{0.2828} & \textbf{0.1125} & \textbf{9.203} \\
 & $\Delta$ & & $+0.185$ & $+0.0057$ & $-0.0066$ & $-0.0019$ & $+0.035$ \\
\midrule
\multirow{4}{*}{RGB $10\%$}
 & Baseline & 15{,}360 & 28.034 & 0.7992 & 0.3292 & 0.1292 & 8.917 \\
 & Baseline$^{+}$ & 16{,}640 & 28.179 & 0.8034 & 0.3241 & 0.1271 & 8.950 \\
 & CVQPG & 15{,}360 & \textbf{28.225} & \textbf{0.8053} & \textbf{0.3217} & \textbf{0.1265} & \textbf{8.966} \\
 & $\Delta$ & & $+0.191$ & $+0.0060$ & $-0.0076$ & $-0.0026$ & $+0.049$ \\
\midrule
\multirow{4}{*}{Grayscale $20\%$}
 & Baseline & 15{,}360 & 29.504 & 0.8340 & 0.3813 & 0.0753 & 8.935 \\
 & Baseline$^{+}$ & 17{,}280 & 29.753 & 0.8394 & 0.3766 & 0.0733 & 8.988 \\
 & CVQPG & 15{,}360 & \textbf{29.834} & \textbf{0.8443} & \textbf{0.3694} & \textbf{0.0725} & \textbf{9.019} \\
 & $\Delta$ & & $+0.331$ & $+0.0103$ & $-0.0119$ & $-0.0028$ & $+0.085$ \\
\midrule
\multirow{4}{*}{Grayscale $10\%$}
 & Baseline & 7{,}680 & 28.076 & 0.7964 & 0.4204 & 0.0888 & 8.548 \\
 & Baseline$^{+}$ & 8{,}640 & 28.317 & 0.8030 & 0.4134 & 0.0862 & 8.621 \\
 & CVQPG & 7{,}680 & \textbf{28.398} & \textbf{0.8079} & \textbf{0.4069} & \textbf{0.0850} & \textbf{8.658} \\
 & $\Delta$ & & $+0.322$ & $+0.0115$ & $-0.0135$ & $-0.0038$ & $+0.109$ \\
\bottomrule
\end{tabular}}
\end{table}

\begin{figure*}[t]
    \centering
    \includegraphics[width=\textwidth]{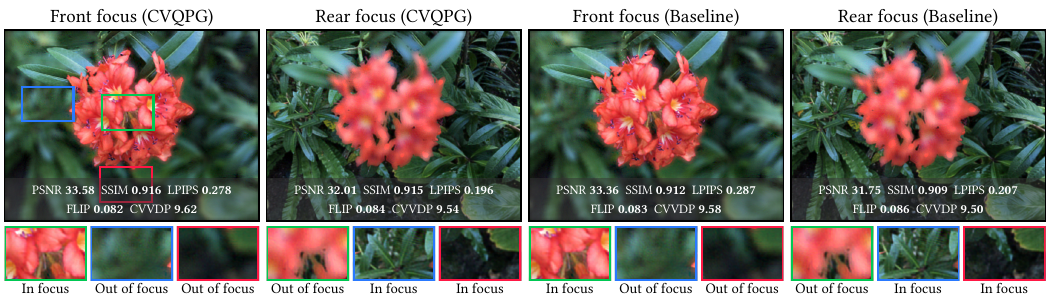}
    \caption{
        Visual comparison on simulated holographic reconstructions (RGB, $20\%$, $N{=}30{,}720$) at the front ($1\,\text{mm}$) and rear ($5\,\text{mm}$) depth planes. CVQPG provides improved visual quality over the baseline. (Source Image: \cite{10.1145/3306346.3322980}) 
    }
    \Description{
        Four photographs of a red flower cluster in front of green foliage, arranged in one row. The left two show the CVQPG reconstruction focused at the front and rear planes; the right two show the baseline reconstruction at the same planes. Below each photograph are five quality metric values and three magnified crops labeled in focus or out of focus. The CVQPG crops show slightly sharper leaf detail and less noise than the baseline crops. 
    }
    \label{fig:results}
\end{figure*}

\begin{figure}[h]
    \centering
    \includegraphics[width=\columnwidth]{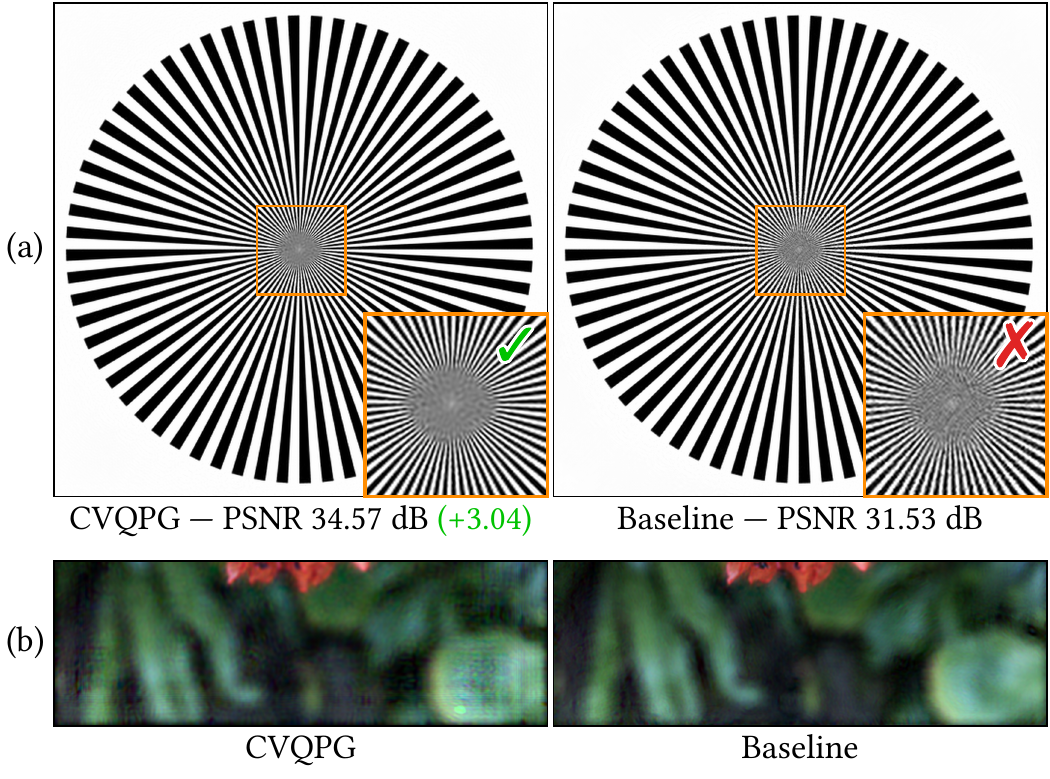}
    \caption{
         Qualitative comparisons. (a) CVQPG outperforms the baseline by $+3.04$\,dB PSNR on the Siemens star target, providing distinct high-frequency details in the central region. (b) Abrupt boundary truncation induces ringing artifacts in CVQPG, which are absent in the baseline. 
    }
    \Description{
        A two-row qualitative comparison. Row (a): two grayscale Siemens star reconstructions side by side, CVQPG at 34.57 dB and the baseline at 31.53 dB, each with a yellow box showing central regions. Row (b): two crops of foliage from a flower reconstruction. CVQPG crop shows ripple-like ringing across the smooth background, whereas the baseline crop is smoother. 
    }
    \label{fig:visual_comparison}
\end{figure}

\paragraph{Qualitative Observations}
To further explore the high-frequency feature of CVQPG, we reconstruct a single-plane, grayscale Siemens star. With equal primitives, CVQPG achieves a significant PSNR gain ($+3.04$\,dB) with a distinct visual improvement in the central region (Fig.~\ref{fig:visual_comparison}a), demonstrating its capacity to resolve high-frequency details. CVQPG exhibits ringing artifacts under zero-padding due to the abrupt truncation of its chirp phase, which is avoided by the baseline (Fig.~\ref{fig:visual_comparison}b). We suppress this by applying a cosine-tapered window to smooth the hologram edges before propagation. 

\paragraph{Ablation study}
The ablation study covers two components: the Scale-Adaptive Curvature Control and the Mahalanobis QPF. Starting from full CVQPG, we compare five scenes at the RGB $10\%$ budget (Tab.~\ref{tab:ablation_results}). Replacing the scale-adaptive bound with a constant bound significantly reduces the PSNR gain ($+0.185 \rightarrow +0.048$\,dB), identifying it as the decisive component. Replacing the Mahalanobis QPF with the standard QPF costs $0.036$\,dB on average, which confirms the benefit of the anisotropic phase profile. Finally, a vanilla model without the two components retains only a gain of $+0.046$\,dB. This performance floor indicates that simply appending a standard QPF fails to capture the full potential of phase modulation. 

\begin{table}[h]
\centering
\caption{
    Component ablation averaged over five scenes (RGB, $10\%$, $N{=}15{,}360$). $\Delta$ denotes the mean PSNR gain over the baseline. Best in \textbf{bold}. 
}
\label{tab:ablation_results}
\resizebox{\linewidth}{!}{%
\begin{tabular}{lcccccc}
\toprule
Variant & $\Delta$ & PSNR$\uparrow$ & SSIM$\uparrow$ & LPIPS$\downarrow$ & FLIP$\downarrow$ & CVVDP$\uparrow$ \\
\midrule
Baseline & --- & 29.809 & 0.8419 & 0.3119 & 0.1128 & 9.095 \\
Full (CVQPG) & $+0.185$ & \textbf{29.994} & \textbf{0.8471} & \textbf{0.3041} & 0.1109 & \textbf{9.137} \\
w/o Adaptive Bound & $+0.048$ & 29.857 & 0.8438 & 0.3103 & 0.1122 & 9.108 \\
w/o Mahalanobis QPF & $+0.149$ & 29.958 & 0.8462 & 0.3044 & \textbf{0.1108} & 9.134 \\
None (Vanilla) & $+0.046$ & 29.855 & 0.8437 & 0.3098 & 0.1118 & 9.109 \\
\bottomrule
\end{tabular}}
\end{table}

\paragraph{Conclusion and Discussion}
CVQPG improves reconstruction quality across all metrics without per-scene tuning and maintains its advantage under equal-parameter evaluation, showing that optimizing wavefronts is more effective than scaling up the count of planar Gaussians. Experiments show that our method achieves its full potential only when the curvature is strictly bounded by the per-primitive sampling limit. We also find that CVQPG preserves mid-to-high frequencies of natural images better than the baseline, making the gain content-dependent. Tab.~\ref{tab:main_results} reveals a chromatic limitation: the grayscale gains notably exceed the RGB ones because a shared curvature cannot be optimal for all wavelengths. In future work, we plan to explore wavelength-based curvature to address the chromatic limitation. We also aim to explore advanced phase formulations and further investigate the representational capacity of explicit hologram representations. 


\bibliographystyle{ACM-Reference-Format}
\bibliography{main}

@inproceedings{zhan2025complexvalued2d,
  author = {Zhan, Yicheng and Gao, Xiangjun and Quan, Long and Ak\c{s}it, Kaan},
  title = {Complex-Valued 2D Gaussian Representation for Computer-Generated Holography},
  booktitle = {Computer Vision -- ECCV 2026},
  month = sep,
  year = {2026},
  publisher = {Springer Nature Switzerland},
  address = {Cham},
  pages = {557--576},
  isbn = {978-3-032-37574-2},
  doi = {10.1007/978-3-032-37574-2_31},
  url = {https://doi.org/10.1007/978-3-032-37574-2_31},
}

@INPROCEEDINGS{10108460,
  author={Kavaklı, Koray and Itoh, Yuta and Urey, Hakan and Akşit, Kaan},
  booktitle={2023 IEEE VR}, 
  title={Realistic Defocus Blur for Multiplane Computer-Generated Holography}, 
  year={2023},
  volume={},
  number={},
  pages={418-426},
  doi={10.1109/VR55154.2023.00057}}

@ARTICLE{10586270,
  author={Xie, Xingyu and Zhou, Pan and Li, Huan and Lin, Zhouchen and Yan, Shuicheng},
  journal={IEEE TPAMI}, 
  title={Adan: Adaptive Nesterov Momentum Algorithm for Faster Optimizing Deep Models}, 
  year={2024},
  volume={46},
  number={12},
  pages={9508-9520},
  doi={10.1109/TPAMI.2024.3423382}}

@INPROCEEDINGS{8578166,
  author={Zhang, Richard and Isola, Phillip and Efros, Alexei A. and Shechtman, Eli and Wang, Oliver},
  booktitle={2018 IEEE/CVF CVPR}, 
  title={The Unreasonable Effectiveness of Deep Features as a Perceptual Metric}, 
  year={2018},
  volume={},
  number={},
  pages={586-595},
  doi={10.1109/CVPR.2018.00068}}

@article{10.1145/3406183,
author = {Andersson, Pontus and Nilsson, Jim and Akenine-M\"{o}ller, Tomas and Oskarsson, Magnus and \r{A}str\"{o}m, Kalle and Fairchild, Mark D.},
title = {FLIP: A Difference Evaluator for Alternating Images},
year = {2020},
issue_date = {August 2020},
publisher = {Association for Computing Machinery},
address = {New York, NY, USA},
volume = {3},
number = {2},
url = {https://doi.org/10.1145/3406183},
doi = {10.1145/3406183},
journal = {Proc. ACM Comput. Graph. Interact. Tech.},
month = aug,
articleno = {15},
numpages = {23}
}

@article{10.1145/3658144,
author = {Mantiuk, Rafal K. and Hanji, Param and Ashraf, Maliha and Asano, Yuta and Chapiro, Alexandre},
title = {ColorVideoVDP: A visual difference predictor for image, video and display distortions},
year = {2024},
issue_date = {July 2024},
publisher = {Association for Computing Machinery},
address = {New York, NY, USA},
volume = {43},
number = {4},
issn = {0730-0301},
url = {https://doi.org/10.1145/3658144},
doi = {10.1145/3658144},
journal = {ACM TOG},
month = jul,
articleno = {129},
numpages = {20}
}

@INPROCEEDINGS{8014884,
  author={Agustsson, Eirikur and Timofte, Radu},
  booktitle={2017 IEEE CVPRW}, 
  title={NTIRE 2017 Challenge on Single Image Super-Resolution: Dataset and Study}, 
  year={2017},
  volume={},
  number={},
  pages={1122-1131},
  doi={10.1109/CVPRW.2017.150}}

@misc{yang2024depthv2,
      title={Depth Anything V2}, 
      author={Lihe Yang and Bingyi Kang and Zilong Huang and Zhen Zhao and Xiaogang Xu and Jiashi Feng and Hengshuang Zhao},
      year={2024},
      eprint={2406.09414},
      archivePrefix={arXiv},
      primaryClass={cs.CV},
}

@article{10.1145/3306346.3322980,
author = {Mildenhall, Ben and Srinivasan, Pratul P. and Ortiz-Cayon, Rodrigo and Kalantari, Nima Khademi and Ramamoorthi, Ravi and Ng, Ren and Kar, Abhishek},
title = {Local light field fusion: practical view synthesis with prescriptive sampling guidelines},
year = {2019},
issue_date = {August 2019},
publisher = {Association for Computing Machinery},
address = {New York, NY, USA},
volume = {38},
number = {4},
issn = {0730-0301},
url = {https://doi.org/10.1145/3306346.3322980},
doi = {10.1145/3306346.3322980},
journal = {ACM TOG},
month = jul,
articleno = {29},
numpages = {14}
}

@article{Zhang:17,
author = {Jingzhao Zhang and Nicolas P\'{e}gard and Jingshan Zhong and Hillel Adesnik and Laura Waller},
journal = {Optica},
number = {10},
pages = {1306--1313},
publisher = {Optica Publishing Group},
title = {3D computer-generated holography by non-convex optimization},
volume = {4},
month = {Oct},
year = {2017},
url = {https://opg.optica.org/optica/abstract.cfm?URI=optica-4-10-1306},
doi = {10.1364/OPTICA.4.001306},
}

@misc{sitzmann2020implicitneuralrepresentationsperiodic,
      title={Implicit Neural Representations with Periodic Activation Functions}, 
      author={Vincent Sitzmann and Julien N. P. Martel and Alexander W. Bergman and David B. Lindell and Gordon Wetzstein},
      year={2020},
      eprint={2006.09661},
      archivePrefix={arXiv},
      primaryClass={cs.CV},
}

@misc{zhang2024gaussianimage1000fpsimage,
      title={GaussianImage: 1000 FPS Image Representation and Compression by 2D Gaussian Splatting}, 
      author={Xinjie Zhang and Xingtong Ge and Tongda Xu and Dailan He and Yan Wang and Hongwei Qin and Guo Lu and Jing Geng and Jun Zhang},
      year={2024},
      eprint={2403.08551},
      archivePrefix={arXiv},
      primaryClass={eess.IV},
}

@misc{windmill2013,
  title        = {Colorful windmills Shenzhen China.},
  howpublished = {\href{https://www.flickr.com/photos/volvob12b/9652416010}{Flickr}},
  journal      = {Flickr},
  publisher    = {CC0 1.0},
  author       = {Bernard Spragg},
  year         = {2013}
}

@article{10.1145/3731163,
author = {Choi, Suyeon and Chao, Brian and Yang, Jacqueline and Gopakumar, Manu and Wetzstein, Gordon},
title = {Gaussian Wave Splatting for Computer-Generated Holography},
year = {2025},
issue_date = {August 2025},
publisher = {Association for Computing Machinery},
address = {New York, NY, USA},
volume = {44},
number = {4},
issn = {0730-0301},
url = {https://doi.org/10.1145/3731163},
doi = {10.1145/3731163},
journal = {ACM TOG},
month = jul,
articleno = {57},
numpages = {13}
}

@article{10.1145/3804450,
author = {Zhan, Yicheng and Shin, Dong-Ha and Baek, Seung-Hwan and Ak\c{s}it, Kaan},
title = {Complex-Valued Holographic Radiance Fields},
year = {2026},
issue_date = {June 2026},
publisher = {Association for Computing Machinery},
address = {New York, NY, USA},
volume = {45},
number = {3},
issn = {0730-0301},
url = {https://doi.org/10.1145/3804450},
doi = {10.1145/3804450},
journal = {ACM TOG},
month = apr,
articleno = {31},
numpages = {16}
}

\end{document}


\title{Hologram Representation via Quadratic Phase Gaussian Splatting}

\fancyhf{}
\renewcommand{\headrulewidth}{0pt}
\renewcommand{\footrulewidth}{0pt}
\pagestyle{fancy}

\maketitle

\thispagestyle{empty}

\section{Holographic Display Prototype}
\begin{figure}[t]
    \centering
    \includegraphics[width=0.95\columnwidth]{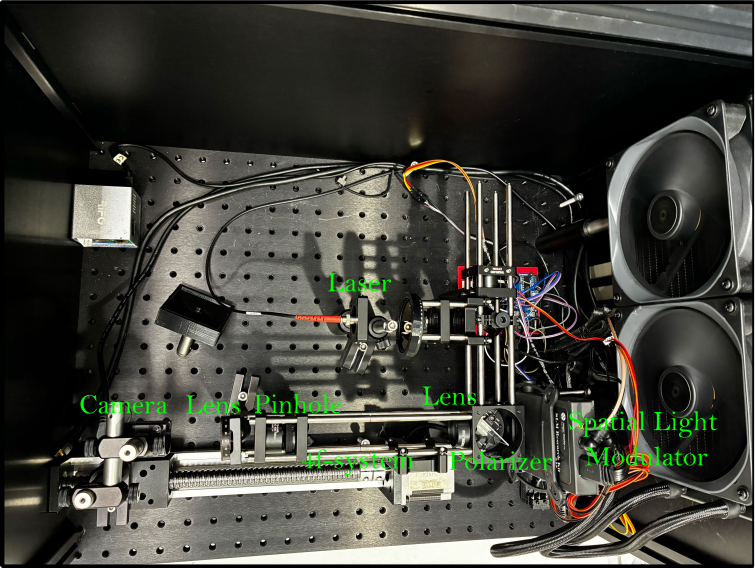}
    \caption{
        Holographic display prototype in the lab. 
    }
    \Description{
        A photograph of the optical bench inside a dark enclosure. Green text labels mark the components along the beam path. A fiber-coupled laser head sits near the center of the bench and emits toward the right, where a polarizer precedes the spatial light modulator, a flat panel with two large cooling fans behind it. The modulated beam returns to the left through a pair of lenses forming the 4f-system, with a pinhole between them, and reaches the camera at the far left, which is mounted on a long motorized linear stage. Cables run from the modulator to the right edge of the enclosure. 
    }
    \label{fig:hardware}
\end{figure}

\begin{figure}[t]
    \centering
    \includegraphics[width=0.95\columnwidth]{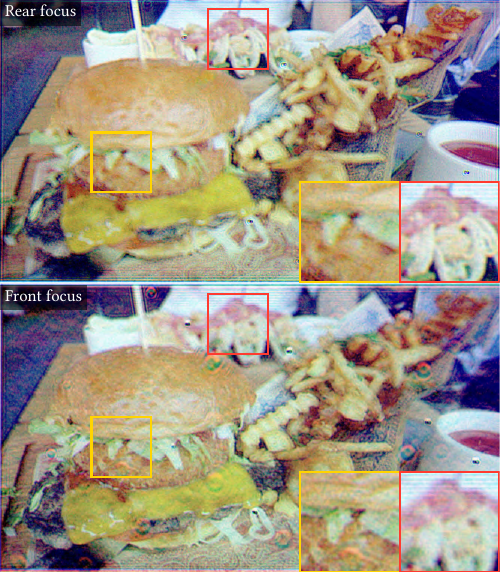}
    \caption{
        Experimentally captured reconstruction result of the Burger scene on our prototype. We display phase-only holograms encoded from the CVQPG representation and photograph the reconstruction at each depth plane. (Source Image: \href{https://openverse.org/image/4a8a7674-7207-4f15-a8ce-c54abe579c5d?q=burger}{kennejima})
    }
    \Description{
        Two photographs of the same reconstructed scene, stacked vertically and labelled Rear focus (upper) and Front focus (lower). Both show a cluster of pink and white rhododendron blossoms surrounded by long green leaves, reproduced from a hologram displayed on a spatial light modulator and recorded directly by a sensor without a lens. 
    }
    \label{fig:capture}
\end{figure}

We employ a holographic display prototype to display reconstructed holograms in our experiments (Fig.~\ref{fig:hardware}). A laser source (LASOS MCS4) combines three laser lines and couples them into a single-mode fiber. A plano-convex lens (Thorlabs LA1708-A, $f=200\,\text{mm}$) collimates the light from the fiber, and a beamsplitter (Thorlabs BP245B1) reflects the collimated, linearly polarized beam onto a phase-only SLM (Jasper JD7714, $4096 \times 2400$ resolution, $3.74\,\mu\text{m}$ pixel pitch). The modulated beam propagates through two lenses (Thorlabs LA1908-A and LB1156-A, $f=500\,\text{mm}$ and $f=250\,\text{mm}$), and a pinhole (Thorlabs SM1D12C) in their common focal plane performs spatial filtering. A lensless image sensor (Point Grey GS3-U3-23S6M-C) mounted on an X-stage (Thorlabs PT1/M) captures the reconstructed holographic images. We further fit CVQPG at $3840 \times 2160$, encode the holograms into phase-only patterns, and demonstrate them on our display prototype. Fig.~\ref{fig:capture} shows an example of the experimentally captured result. 

\section{Runtime and Memory Consumption}
We profile the training runtime and peak memory consumption of the PyTorch pipelines. CVQPG trains in $8.7$--$26.5$\,min using $5.5$--$18.9$\,GiB of memory, adding at most $3.4\%$ time and $5.9\%$ memory overhead over the Baseline at equal primitive counts (Tab.~\ref{tab:runtime}). Under equal-parameter evaluation, CVQPG trains $3.4$--$5.4\%$ faster and uses $3.8$--$6.8\%$ less memory than Baseline$^{+}$ while achieving higher reconstruction quality. 

\begin{table}[h]
\centering
\caption{
    Training time and peak GPU memory averaged over the ten scenes (PyTorch pipeline, 2000 steps, $640\times480$, two depth planes, a single NVIDIA TITAN RTX). \#Params counts the trainable parameters. $\Delta$ columns show the relative change with respect to the standard Baseline. 
}
\label{tab:runtime}
\resizebox{\linewidth}{!}{%
\begin{tabular}{llcccccc}
\toprule
Setting & Method & $N$ & \#Params & Time (min) & $\Delta$Time & Memory (GiB) & $\Delta$Memory \\
\midrule
\multirow{3}{*}{RGB $20\%$}
 & Baseline & 30{,}720 & 368{,}640 & 26.6 & --- & 18.9 & --- \\
 & Baseline$^{+}$ & 33{,}280 & 399{,}360 & 28.1 & $+5.7\%$ & 20.3 & $+7.2\%$ \\
 & CVQPG & 30{,}720 & 399{,}360 & 26.5 & $0.0\%$ & 18.9 & $-0.1\%$ \\
\midrule
\multirow{3}{*}{RGB $10\%$}
 & Baseline & 15{,}360 & 184{,}320 & 17.0 & --- & 10.9 & --- \\
 & Baseline$^{+}$ & 16{,}640 & 199{,}680 & 17.8 & $+4.5\%$ & 11.6 & $+5.7\%$ \\
 & CVQPG & 15{,}360 & 199{,}680 & 17.0 & $0.0\%$ & 10.9 & $-0.1\%$ \\
\midrule
\multirow{3}{*}{Grayscale $20\%$}
 & Baseline & 15{,}360 & 122{,}880 & 13.0 & --- & 9.0 & --- \\
 & Baseline$^{+}$ & 17{,}280 & 138{,}240 & 14.1 & $+8.2\%$ & 9.9 & $+10.3\%$ \\
 & CVQPG & 15{,}360 & 138{,}240 & 13.4 & $+3.0\%$ & 9.5 & $+5.9\%$ \\
\midrule
\multirow{3}{*}{Grayscale $10\%$}
 & Baseline & 7{,}680 & 61{,}440 & 8.4 & --- & 5.2 & --- \\
 & Baseline$^{+}$ & 8{,}640 & 69{,}120 & 9.0 & $+7.1\%$ & 5.7 & $+9.4\%$ \\
 & CVQPG & 7{,}680 & 69{,}120 & 8.7 & $+3.4\%$ & 5.5 & $+5.2\%$ \\
\bottomrule
\end{tabular}}
\end{table}

\section{Per-Scene Results}
Tab.~\ref{tab:per_scene} reports the results of all three models on every scene and setting. CVQPG improves upon the Baseline in all $200$ comparisons and upon the Baseline$^{+}$ in $182$ of them. The results show that CVQPG provides consistent improvements at equal primitive counts, and these gains are unbiased across the dataset. 

\begin{table*}[h]
\centering
\small
\setlength{\tabcolsep}{3.2pt}
\renewcommand{\arraystretch}{0.92}
\caption{
    Per-scene results for all four settings, averaged over the two depth planes. Each cell lists Baseline\,/\,Baseline$^{+}$\,/\,CVQPG. Our approach CVQPG is the best of the three in 182 of the 200 cells. 
}
\label{tab:per_scene}
\resizebox{\linewidth}{!}{%
\begin{tabular}{ll ccccc ccccc}
\toprule
 & & \multicolumn{5}{c}{RGB} & \multicolumn{5}{c}{Grayscale} \\
\cmidrule(lr){3-7}\cmidrule(lr){8-12}
Scene & Rate & PSNR$\uparrow$ & SSIM$\uparrow$ & LPIPS$\downarrow$ & FLIP$\downarrow$ & CVVDP$\uparrow$ & PSNR$\uparrow$ & SSIM$\uparrow$ & LPIPS$\downarrow$ & FLIP$\downarrow$ & CVVDP$\uparrow$ \\
\midrule
\multirow{2}{*}{Pencil}
 & $20\%$ & 26.967/27.094/27.115 & .7737/.7766/.7792 & .2946/.2893/.2876 & .1528/.1516/.1508 & 8.749/8.770/8.780 & 26.788/27.032/27.074 & .7738/.7820/.7861 & .3829/.3783/.3690 & .0856/.0826/.0831 & 8.592/8.673/8.683 \\
 & $10\%$ & 25.798/25.935/25.931 & .7391/.7441/.7434 & .3295/.3284/.3239 & .1700/.1660/.1684 & 8.488/8.534/8.520 & 25.364/25.602/25.557 & .7319/.7365/.7403 & .4268/.4203/.4118 & .1006/.0991/.0982 & 8.176/8.243/8.256 \\
\addlinespace[2pt]
\multirow{2}{*}{Sushi}
 & $20\%$ & 33.646/33.766/33.904 & .9199/.9222/.9242 & .2555/.2542/.2504 & .0826/.0817/.0804 & 9.511/9.519/9.536 & 34.382/34.647/34.843 & .9286/.9323/.9355 & .3227/.3202/.3132 & .0518/.0504/.0492 & 9.412/9.447/9.483 \\
 & $10\%$ & 32.364/32.527/32.589 & .8955/.8987/.9002 & .2888/.2848/.2836 & .0935/.0916/.0913 & 9.347/9.371/9.381 & 32.928/33.122/33.369 & .9023/.9077/.9116 & .3577/.3506/.3455 & .0609/.0589/.0574 & 9.133/9.186/9.228 \\
\addlinespace[2pt]
\multirow{2}{*}{Flower}
 & $20\%$ & 32.554/32.682/32.793 & .9105/.9131/.9158 & .2469/.2409/.2373 & .0843/.0839/.0826 & 9.540/9.555/9.579 & 33.111/33.362/33.522 & .9190/.9237/.9273 & .3204/.3167/.3095 & .0642/.0631/.0620 & 9.417/9.451/9.491 \\
 & $10\%$ & 31.179/31.307/31.413 & .8797/.8831/.8858 & .2916/.2868/.2834 & .0959/.0945/.0939 & 9.285/9.313/9.340 & 31.523/31.807/31.945 & .8820/.8895/.8946 & .3643/.3561/.3477 & .0744/.0722/.0714 & 9.054/9.126/9.174 \\
\addlinespace[2pt]
\multirow{2}{*}{Straw}
 & $20\%$ & 32.322/32.457/32.531 & .9018/.9041/.9062 & .2306/.2282/.2263 & .0926/.0919/.0914 & 9.352/9.376/9.371 & 32.696/32.976/33.074 & .9097/.9143/.9173 & .3415/.3354/.3293 & .0604/.0589/.0584 & 9.208/9.254/9.286 \\
 & $10\%$ & 31.082/31.214/31.310 & .8777/.8805/.8829 & .2640/.2599/.2570 & .1011/.1003/.0989 & 9.243/9.236/9.279 & 31.023/31.335/31.387 & .8770/.8832/.8866 & .3760/.3689/.3651 & .0716/.0691/.0685 & 8.826/8.904/8.935 \\
\addlinespace[2pt]
\multirow{2}{*}{Dragon}
 & $20\%$ & 27.448/27.560/27.604 & .8306/.8337/.8363 & .2916/.2865/.2854 & .1395/.1389/.1379 & 9.071/9.093/9.103 & 27.873/28.134/28.139 & .8386/.8435/.8476 & .3801/.3773/.3716 & .0883/.0866/.0860 & 8.802/8.849/8.869 \\
 & $10\%$ & 26.152/26.330/26.328 & .7952/.8002/.8013 & .3273/.3206/.3220 & .1566/.1543/.1534 & 8.810/8.856/8.856 & 26.282/26.565/26.580 & .7922/.8009/.8045 & .4182/.4101/.4049 & .1051/.1013/.1005 & 8.375/8.455/8.479 \\
\addlinespace[2pt]
\multirow{2}{*}{Tiger}
 & $20\%$ & 28.991/29.096/29.141 & .8549/.8568/.8599 & .3008/.2967/.2930 & .1087/.1081/.1071 & 9.252/9.271/9.283 & 29.124/29.332/29.377 & .8574/.8625/.8670 & .3840/.3801/.3719 & .0732/.0715/.0712 & 9.015/9.069/9.098 \\
 & $10\%$ & 27.968/28.070/28.109 & .8272/.8314/.8330 & .3391/.3325/.3280 & .1205/.1189/.1184 & 8.996/9.032/9.042 & 28.027/28.201/28.303 & .8278/.8335/.8385 & .4165/.4116/.4049 & .0830/.0810/.0799 & 8.693/8.756/8.797 \\
\addlinespace[2pt]
\multirow{2}{*}{Windmill}
 & $20\%$ & 24.709/24.840/24.886 & .6859/.6902/.6956 & .3326/.3290/.3259 & .1388/.1371/.1361 & 8.826/8.851/8.880 & 24.995/25.244/25.284 & .6784/.6874/.6965 & .4680/.4604/.4528 & .0952/.0920/.0914 & 8.348/8.414/8.462 \\
 & $10\%$ & 23.441/23.579/23.650 & .6450/.6493/.6551 & .3702/.3658/.3615 & .1590/.1571/.1556 & 8.509/8.547/8.579 & 23.785/23.987/24.099 & .6339/.6407/.6504 & .5032/.4978/.4895 & .1107/.1075/.1056 & 7.909/7.993/8.043 \\
\addlinespace[2pt]
\multirow{2}{*}{Statue}
 & $20\%$ & 23.586/23.732/23.749 & .6556/.6605/.6652 & .3784/.3699/.3701 & .1561/.1543/.1533 & 8.634/8.670/8.709 & 23.398/23.619/23.729 & .6506/.6562/.6677 & .5037/.4969/.4867 & .1055/.1017/.0993 & 8.243/8.327/8.393 \\
 & $10\%$ & 22.401/22.538/22.588 & .6136/.6184/.6231 & .4375/.4304/.4302 & .1835/.1800/.1778 & 8.225/8.279/8.334 & 22.215/22.361/22.454 & .6022/.6096/.6180 & .5598/.5513/.5428 & .1285/.1248/.1219 & 7.669/7.767/7.830 \\
\addlinespace[2pt]
\multirow{2}{*}{Burger}
 & $20\%$ & 32.860/32.940/33.030 & .8947/.8966/.8985 & .2758/.2736/.2700 & .0745/.0740/.0730 & 9.576/9.583/9.600 & 32.967/33.185/33.306 & .8993/.9035/.9070 & .3531/.3494/.3447 & .0545/.0528/.0521 & 9.413/9.450/9.482 \\
 & $10\%$ & 31.738/31.890/31.931 & .8679/.8716/.8730 & .3106/.3074/.3014 & .0843/.0824/.0826 & 9.358/9.393/9.403 & 31.687/31.889/32.044 & .8689/.8746/.8794 & .3794/.3724/.3668 & .0636/.0615/.0603 & 9.101/9.168/9.214 \\
\addlinespace[2pt]
\multirow{2}{*}{Bento}
 & $20\%$ & 29.567/29.693/29.746 & .8797/.8822/.8832 & .2879/.2850/.2823 & .1142/.1125/.1123 & 9.166/9.168/9.189 & 29.703/29.997/29.995 & .8841/.8886/.8908 & .3567/.3513/.3453 & .0746/.0734/.0720 & 8.899/8.946/8.947 \\
 & $10\%$ & 28.218/28.403/28.401 & .8515/.8570/.8549 & .3336/.3243/.3257 & .1275/.1255/.1252 & 8.907/8.935/8.926 & 27.925/28.302/28.242 & .8463/.8536/.8552 & .4025/.3953/.3901 & .0897/.0862/.0862 & 8.545/8.614/8.620 \\
\midrule
\multirow{2}{*}{Mean}
 & $20\%$ & 29.265/29.386/29.450 & .8307/.8336/.8364 & .2895/.2853/.2828 & .1144/.1134/.1125 & 9.168/9.186/9.203 & 29.504/29.753/29.834 & .8340/.8394/.8443 & .3813/.3766/.3694 & .0753/.0733/.0725 & 8.935/8.988/9.019 \\
 & $10\%$ & 28.034/28.179/28.225 & .7992/.8034/.8053 & .3292/.3241/.3217 & .1292/.1271/.1265 & 8.917/8.950/8.966 & 28.076/28.317/28.398 & .7964/.8030/.8079 & .4204/.4134/.4069 & .0888/.0862/.0850 & 8.548/8.621/8.658 \\
\bottomrule
\end{tabular}}
\end{table*}

\section{Visual Results}
\begin{figure*}[h]
    \centering
    \includegraphics[width=\linewidth]{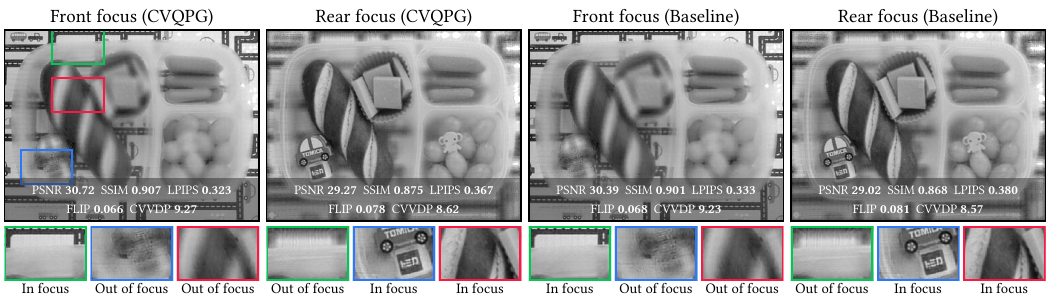}
    \caption{
        Visual comparison on simulated holographic reconstructions between CVQPG and the baseline. We list two representative scenes of RGB $20\%$ and grayscale $20\%$. (Source Image: \href{https://openverse.org/image/90ca6d2b-8c24-48d5-abf2-6f0dc2d81808?q=sushi&p=10}{Ben Sutherland}, \href{https://openverse.org/image/a1c7d020-1d46-4193-8708-a959c5befdb4?q=bento&p=119}{anotherlunch.com})
    }
    \Description{
        Two rows of photographs, each row shows four reconstructed results of a scene. The left two show the CVQPG reconstruction focused at the front and rear planes; the right two show the baseline reconstruction at the same planes. Below each photograph are five quality metric values and three magnified crops labeled in focus or out of focus. 
    }
    \label{fig:demo}
\end{figure*}

Fig.~\ref{fig:demo} illustrates two representative reconstructions from our dataset. 

\begin{figure*}[h]
    \centering
    \includegraphics[width=\linewidth]{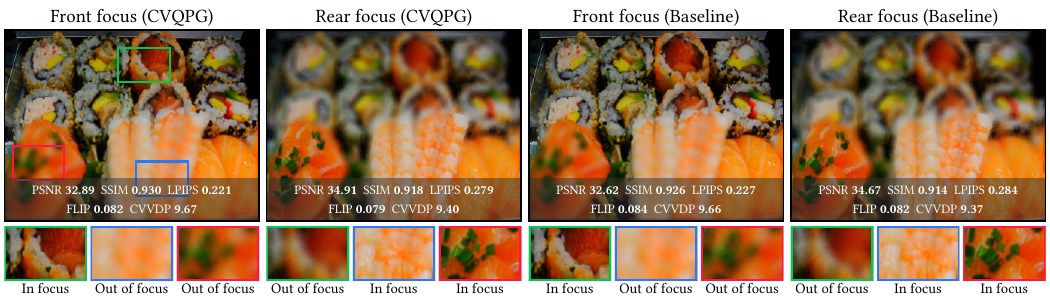}
\end{figure*}

